\documentclass[aps,superscriptaddress,twocolumn]{revtex4}

\usepackage[]{graphicx}

\graphicspath{{./}{fig/}}

\begin{document}

\title{A symmetry breaking mechanism for epithelial cell polarization}

\author{A. Veglio}

\affiliation{Department of Oncological Sciences and Division of Molecular Angiogenesis, Institute for Cancer Research and Treatment, University of Torino School of Medicine, Candiolo (TO), Italy.}

\affiliation{CNISM, Corso Duca degli Abruzzi 24, 10125 Torino, Italy.}

\author{A. Gamba}


\affiliation{Politecnico di Torino, Corso Duca degli Abruzzi 24, 10125 Torino, Italy; and INFN Torino, Italy.}

\affiliation{CNISM, Corso Duca degli Abruzzi 24, 10125 Torino, Italy.}

\author{M. Nicodemi}

\affiliation{Dept. of Physics \& Complexity Sciences, University of Warwick, UK; and INFN Napoli, Italy.}

\author{F. Bussolino}

\affiliation{Department of Oncological Sciences and Division of Molecular Angiogenesis, Institute for Cancer Research and Treatment, University of Torino School of Medicine, Candiolo (TO), Italy.}

\author{G. Serini}

\affiliation{Department of Oncological Sciences and Division of Molecular Angiogenesis, Institute for Cancer Research and Treatment, University of Torino School of Medicine, Candiolo (TO), Italy.}

\begin{abstract}
In multicellular organisms, epithelial cells form layers separating compartments responsible for different physiological functions. 
At the early stage of epithelial layer formation, each cell of an
aggregate defines an inner and an outer side by breaking the symmetry of its initial state, in a process known as epithelial polarization. 
By integrating recent biochemical and biophysical data with stochastic
simulations of the relevant reaction-diffusion system we provide evidence
that epithelial cell polarization is a chemical phase separation process
induced by a local bistability in the signaling network
at the level of the cell membrane.
The 
early symmetry breaking event 
triggering phase separation 
is  
induced
by adhesion-dependent mechanical forces localized in the point of
convergence of cell surfaces when a threshold number of confluent cells is
reached.
The generality of the emerging phase separation scenario is likely common
to many processes of cell polarity formation.
\end{abstract}

\maketitle

\section{Introduction}\label{sec:intro}
\begin{figure}
\centering
\includegraphics[width=8.cm]{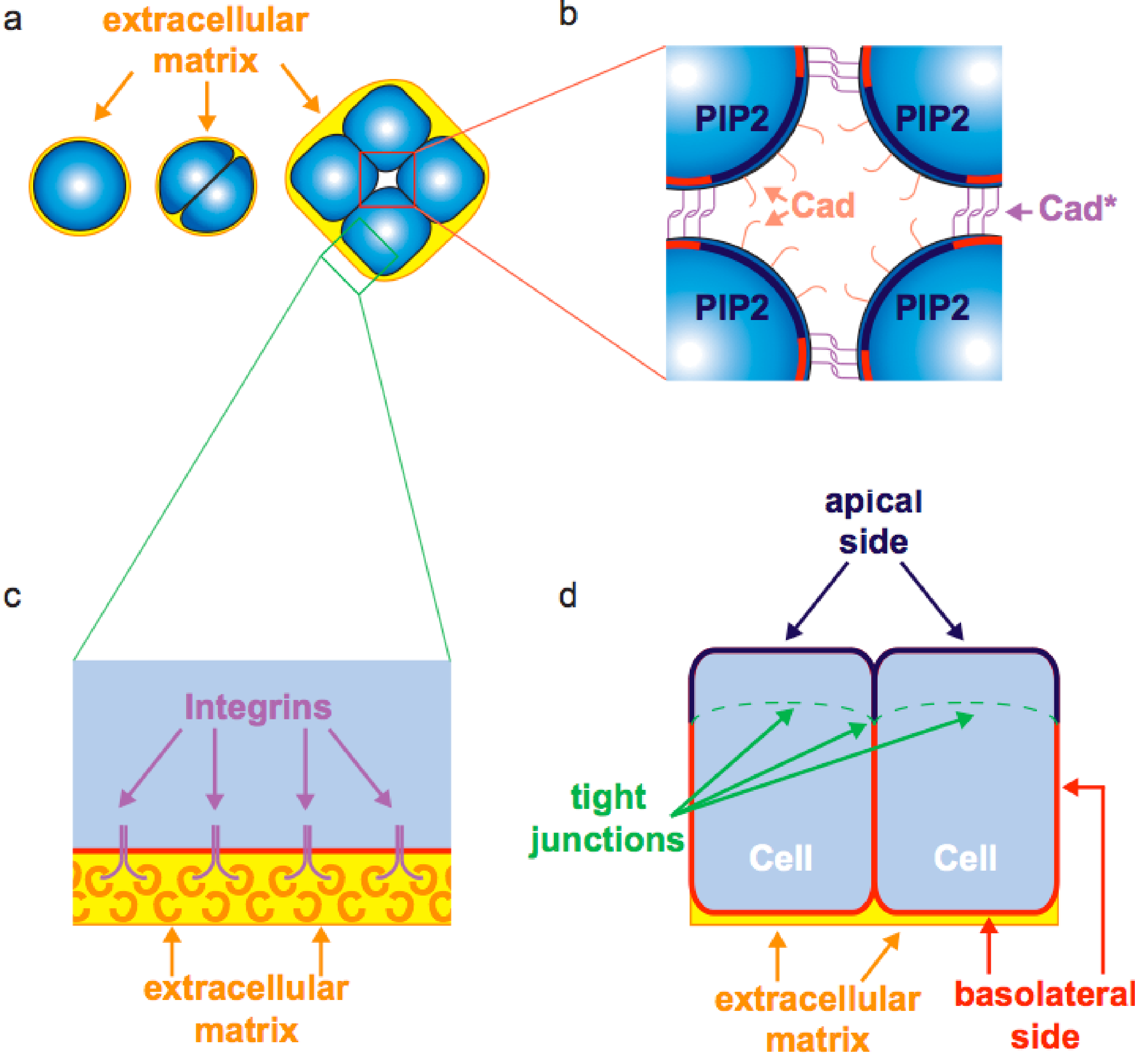}
\caption{(Color online) a,b) After a few cell divisions, at the common point of contact among the cells of an aggregate 
the intrinsic high membrane curvature locally induces formation of a
microlumen (inner circle) and the breaking of cell-cell adhesive receptor
links (pink and purple bars). This breaking
favors the appearance of a small region rich in PIP2 lipids (blue) which provides a nucleation center for the ensuing symmetry breaking process. 
{b,c) Everywhere but in correspondence of the PIP2-rich region, the cell is surrounded by adhesive contacts, either cell-cell (cadherins) or cell-matrix contacts (integrins).}
Cad (pink bars) and Cad$^\ast$ (purple bars) respectively indicate
inactive and active cadherins.
d) Scheme of the 
apical (blue) and basolateral (red) regions 
and the ring of tight junctions 
(green)
spontaneously assembling at their interface. 
The extracellular matrix (yellow) is indicated as well.
}
\label{layer}
\end{figure}

The development of epithelial tissues (e.g.,
kidney tubules, respiratory and gastrointestinal tracts, etc.) results from complex morphogenetic processes implying the arrangement of cells in layers organized along specific directional axes~\cite{Comer07,MGD+07}. Epithelial cells are endowed with a self-polarization mechanism defining an `inner' and `outer' side, which is mandatory to allow organs to exert their vital functions. 
In a well established \emph{in~vitro} cell system, which recapitulates the \emph{in~vivo}
morphogenesis, after a single epithelial cell is seeded in a three-dimensional gel (Fig.~\ref{layer}a), cell division begins, and a multicellular aggregate arises~\cite{Comer07}. The cells in the aggregate are bound 
each other (through cadherin molecules, Fig.~\ref{layer}b) and 
to an {extracellular} matrix 
(through integrin molecules, Fig.~\ref{layer}c).
When the cell
number reaches 5-6 units, an inner cavity, named lumen, is spontaneously opened~\cite{WON90} (Fig.~\ref{layer}a,b).
Afterwards, cells develop a top (called apical) and a bottom (basolateral) side (Fig.~\ref{layer}d) having different chemical features, while cell-cell 
and cell-matrix
contacts only persist in the basolateral region. 
Finally, the border between apical and basolateral sides is sealed by ring-shaped tight junction proteins, which spontaneously find their functional position (Fig.~\ref{layer}d) and prevent intermixing of chemical components between the apical and the basolateral membranes, as well as the outpouring of liquids from the lumen.  
{ 
The full polarization process has a complex nature and involves different 
factors and stages. 
However, recent experiments have determined its 
master 
regulator 
(see \cite{bryant2008,mellman2008} and ref.s therein): intracellular
asymmetry and lumen opening are controlled by PIP2/3 phospholipids (see
Fig.\ref{net}) and their interaction with PTEN/PI3K enzymes which induce
PIP2/3 segregation to opposite poles, while the PAR complex further
stabilizes axial polarity (see the review~\cite{Comer07} and ref.s below). 
Even in such an {\em in vitro} system, however, the mechanism whereby the cell original spatial symmetry is spontaneously broken and polarization develops remains mysterious~\cite{Comer07}. 
}

To describe cell differentiation, polarization and signal localization 
mechanisms, 
stochastic 
reaction-diffusion~\cite{Kam07} and coupled kinetic rate 
equations have been 
widely 
used
for
intracellular signaling, 
gene regulation
and autocatalytic reaction systems 
(see \textit{e.g.}
\cite{Shraiman05,Thattai02,PBE00}
and ref.s therein).
In 
this
context, here we use reaction-diffusion equations to model the PIP2/3 master regulator 
in order to address some 
open questions~\cite{Comer07}, 
namely the mechanisms for: \emph{i)} the lumen site choice; 
\emph{ii)}  its opening; \emph{iii)} the control of its final size;
\emph{iv)} the localization of tight junctions. 
After delineating the chemical reactions involved in the process, by simulation of its
master equation we show that self-polarization can be understood in terms of statistical 
physics concepts as a symmetry breaking mechanism driven by the chemical regulatory network.
We finally interpret the simulation results in terms of a simple mean-field
model.

\section{Model}\label{sec:model}
\begin{figure}
\centering
\includegraphics[width=7cm]{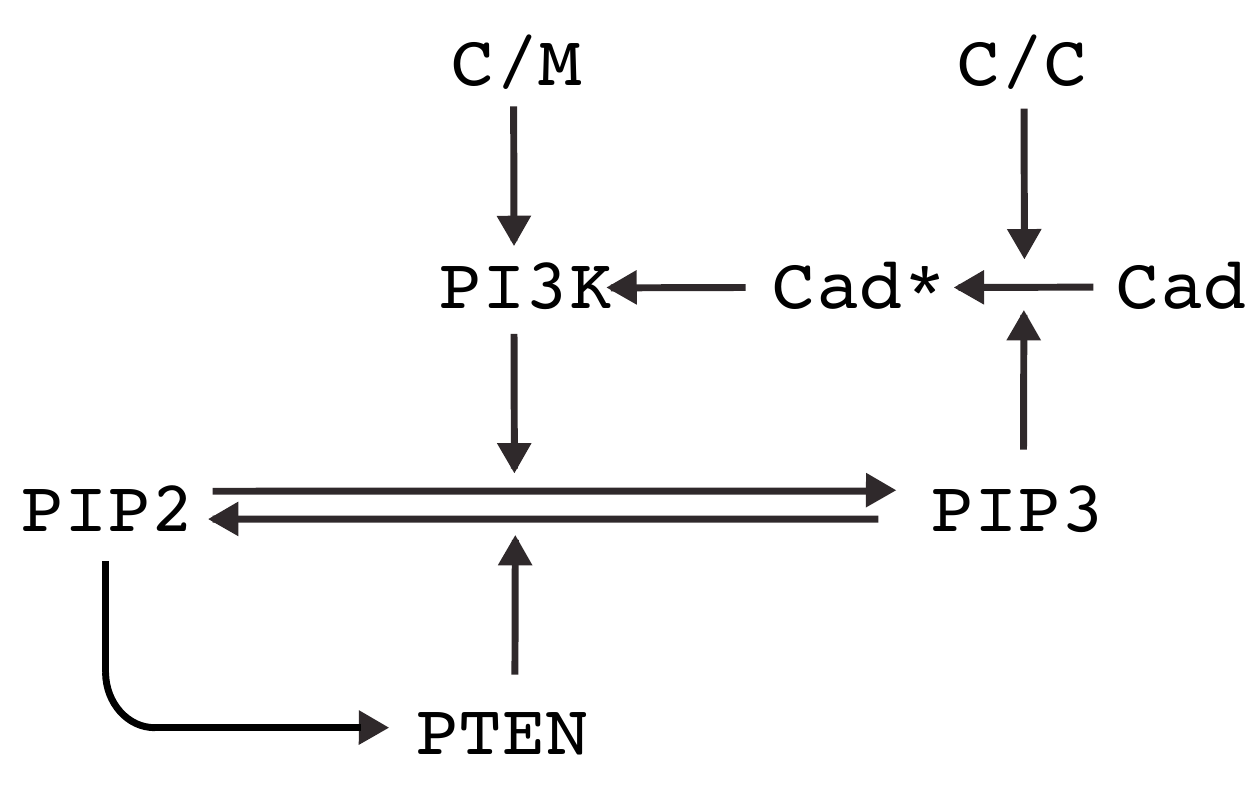}
\caption{Reactions scheme of the epithelial polarization network. The core players are the PI3K and PTEN enzymes, which respectively catalyze the switch of the PIP2 lipid into PIP3 and \emph{vice versa}. 
{PTEN becomes effective upon association to PIP2, while}  
PI3K becomes effective  upon association with activated
cell-cell adhesive receptors (cadherins, Cad$^{*}$) {or }{extracellular} matrix adhesive receptors (C/M). To bind PI3K, cadherins (Cad) must be activated (Cad$^{\ast}$) by linking other cadherins on neighboring cells (C/C) and by {the} action of PIP3. 
The symmetry of PTEN/PI3K in {generating} PIP2/3 can be broken by cadherins inactivation. {The PIP2-PTEN feedback loop may then locally prevail, originating the chemical polarization of the cell.} 
}
\label{net}
\end{figure}

Recent 
experiments
have shown that 
PIP2
and 
PIP3
for a module that
acts as a
master regulator 
controlling all signaling pathways and cytoskeletal
dynamics required for epithelial cell 
polarization~\cite{bryant2008,mellman2008}. 
PIP2/PIP3 levels are regulated by
the counteracting enzymes 
PI3K
and 
PTEN, which
respectively catalyze the switch of PIP2 to PIP3 and 
\emph{vice versa}~\cite{Comer07,KCF07} (Fig.~\ref{net}).
The phospholipids (PIP2/3) are stably localized in the inner face of the
cell membrane where they diffuse. 
The enzymes (PI3K/PTEN) diffuse 
instead
in
the cell volume, where they are present in limiting amounts, {and} become
active upon association with membrane spanning proteins or lipids. 
PTEN levels in the membrane are controlled by its binding to PIP2,
thus 
realizing
a positive feedback loop
(see Fig.~\ref{net}).
PI3K levels in the membrane are controlled by its binding to cell-cell 
adhesive receptors (cadherins) and cell matrix adhesive receptors 
(integrins, schematically indicated by C/M in Fig. \ref{net})~\cite{WD99}. 
To bind PI3K, cadherins must be \emph{activated} 
(Cad$^{\ast}$ in Fig.~\ref{layer}b and \ref{net}) 
by engagement with cadherins of a neighboring cell 
(named C/C in the diagram of 
Fig.~\ref{net})
PI3K is active 
only when associated to either activated cadherins or integrins. 
Since PIP3 stabilizes the activated form Cad$^{\ast}$~\cite{YK03}, 
these interactions create a positive PIP3-PI3K feedback loop, mediated by the existence of cell-cell contacts (Fig.~\ref{net}).
Before polarization, cadherins and integrins are activated along the whole membrane and PIP3 uniformly prevails on PIP2 determining a stable PIP3-rich phase over the whole membrane.
A local depletion of PIP3-PI3K can be created if a large enough membrane area with disrupted cell-cell links is formed, {thereby} breaking the PIP3-PI3K feedback loop (Fig.~\ref{net}) and originating a germ of a PIP2-rich phase
(Fig.~\ref{layer}b and~\ref{net}). 
Then, the PIP2-PTEN feedback loop may locally prevail, inducing a
PIP2 and PIP3 surface compartmentalization that splits the cell membrane in two {regions, or phases~\cite{LL80}, characterized by different chemical concentrations of the signaling molecules}.

Several mechanisms have been proposed~\cite{bryant2008} to explain the
initiation of the polarization process, which is typically found in high
curvature membrane regions, especially in correspondence of multiple
cell-cell contacts (cellular vertex)~\cite{MGD+07} away from the extracellular matrix (Fig.~\ref{layer}). 
We observe here that
cell-cell contacts can be broken~\cite{Kro04} 
as soon as the adhesion energy $W_{\mathrm{a}}$ (\textit{i.e.} the energy per unit area
needed to break cadherin contacts) 
becomes comparable to the elastic energy stored in the cell membrane in high
curvature regions, such as at the confluence of several cells
(Fig.~\ref{layer}a,b).
Since the elastic energy per unit area $W_{\mathrm{e}}$ stored in the membrane is~\cite{LL86, Fou07, SWF+98}
\begin{equation}
W_\mathrm{e} = \kappa/2r^2,
\label{wakappe}
\end{equation} 
where $\kappa$ is the membrane bending rigidity and $r$ the local curvature radius~\cite{Fou07}, the condition $W_{\mathrm{a}}\sim W_{\mathrm{e}}$ allows to estimate the critical curvature radius $r_\mathrm{a}$ where cell-cell contacts start being disrupted as~\cite{SWF+98} 
\begin{equation}
r_\mathrm{a}\simeq\sqrt{\kappa/2 W_\mathrm{a}}\ .
\label{wakappa}
\end{equation} 
The critical value 
$r_\mathrm{a}$
can be easily estimated.
In eukaryotic cells,
the typical adhesion energy of cadherin contacts is
$W_\mathrm{a} \simeq 10^{- 11} \mu \mathrm{J} / \mu
\mathrm{m}^2$~\cite{SWF+98},
while the typical bending rigidity is
$\kappa \simeq 400\, k_\mathrm{B} T \simeq 16 \times 10^{- 13} \mu
\mathrm{J}$~\cite{SWF+98}, 
giving~\footnote{It is worth observing here that 
since the values $\kappa$ and
$W_\mathrm{a}$ 
used for the estimate
have been
measured on live cells~\cite{SWF+98}, 
they take effectively 
into account the mechanical contributions from both the lipid bilayer and
the cortical cytoskeleton. }
\begin{equation}
r_\mathrm{a} \simeq 0.3\, \mu \mathrm{m}\ .
\label{defra}
\end{equation}

According to the above estimate,
when the number of cells in the initial aggregate increases up to the 5-6
cell stage, at 
the cell convergence
points (see Fig.~\ref{layer}b) the membrane curvature increases {as well} and, especially in areas not in contact with the 
extracellular matrix, cadherin bridges are subject to forces that can disrupt links.
By such a mechanism, a local opening of cell-cell contacts breaks the
PIP3-PI3K feedback loop and induces a local unbalance towards PIP2
formation (Fig.~\ref{net}) and a germ of the PIP2{-rich} phase {can be} 
ushered in (Fig.~\ref{layer}b). 


In Sect.~\ref{sec:simul} we show that only germs of the PIP2-rich phase larger than a treshold radius $r_{\mathrm{thr}}$ actually survive and grow.
This fact suggests that although the uniform PIP3-rich phase is not the more
stable state for the signaling network, a polarized state characterized by
the coexistence of the PIP3-rich and the PIP2-rich phase may be reached only
by overcoming a barrier in a suitably defined effective energy (see 
Sect.~\ref{sec:mf} for a detailed discussion of this point).
Therefore, we are faced with the following physical picture: if elastic forces due to high membrane curvature in the region of cell convergence (cellular vertex) trigger disruption of cadherin links in a region of size larger than $r_{\mathrm{thr}}$, the PIP2-rich patch grows favoring further breaking of cadherin links (Fig.~\ref{net}) and the formation of a lumen.
Thus, for the process of lumen formation to start, it is necessary that the local curvature radius in the cellular vertex satisfies
\begin{equation}
r_{\mathrm{a}}\gtrsim r_{\mathrm{thr}}\ .
\end{equation} 

The growth of the PIP2-patch and lumen slows down and eventually comes to a stop as soon as cytosolic PTEN is depleted.
This way, at the end of the process the cell reaches a stable polarized state characterized by the coexistence of the PIP2-rich and the PIP3-rich phase~\cite{FCG+08, GKL+07}, and a lumen coinciding with the PIP2-rich phase is formed.

\section{Simulations.}\label{sec:simul}
In this Section we investigate on quantitative grounds {the above described scenario} 
of polarization. 

Since the chemical reaction and diffusion processes are intrisically noisy, we simulate the corresponding dynamics by a stochastic algorithm, using realistic values for reaction and diffusion rates.
We can check this way that noise alone is not sufficient here to overcome the energy barrier separating the uniform and polarized state in observational times, if an initial PIP2 seed of size larger than $r_{\mathrm{thr}}$ is not created by an external interaction.

\begin{figure}[h]
  \begin{tabular}{lccc}
  \ &&&\\
  \resizebox{1.4cm}{!}{\includegraphics{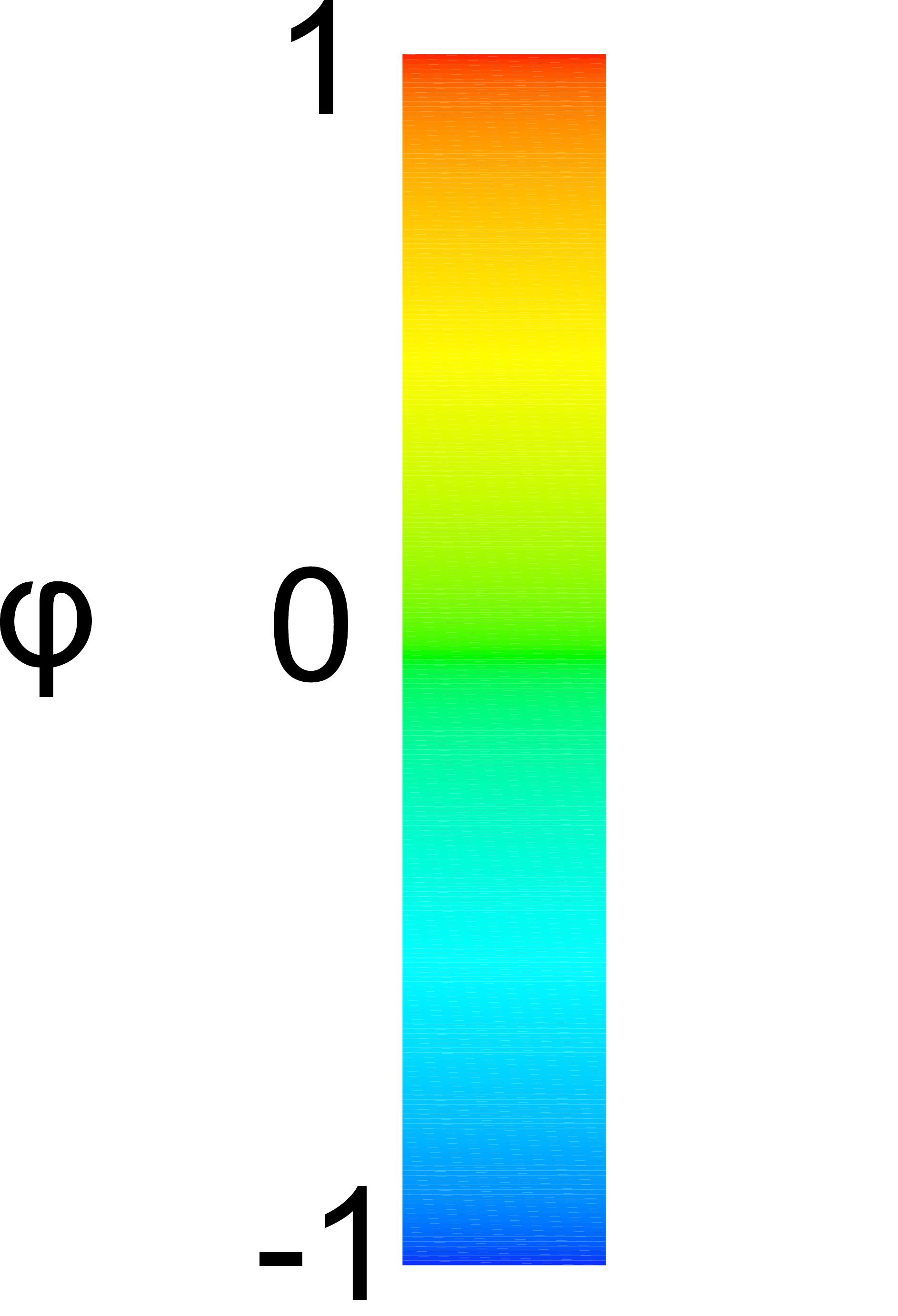}}\ \ \ &
  \resizebox{2.2cm}{!}{\includegraphics{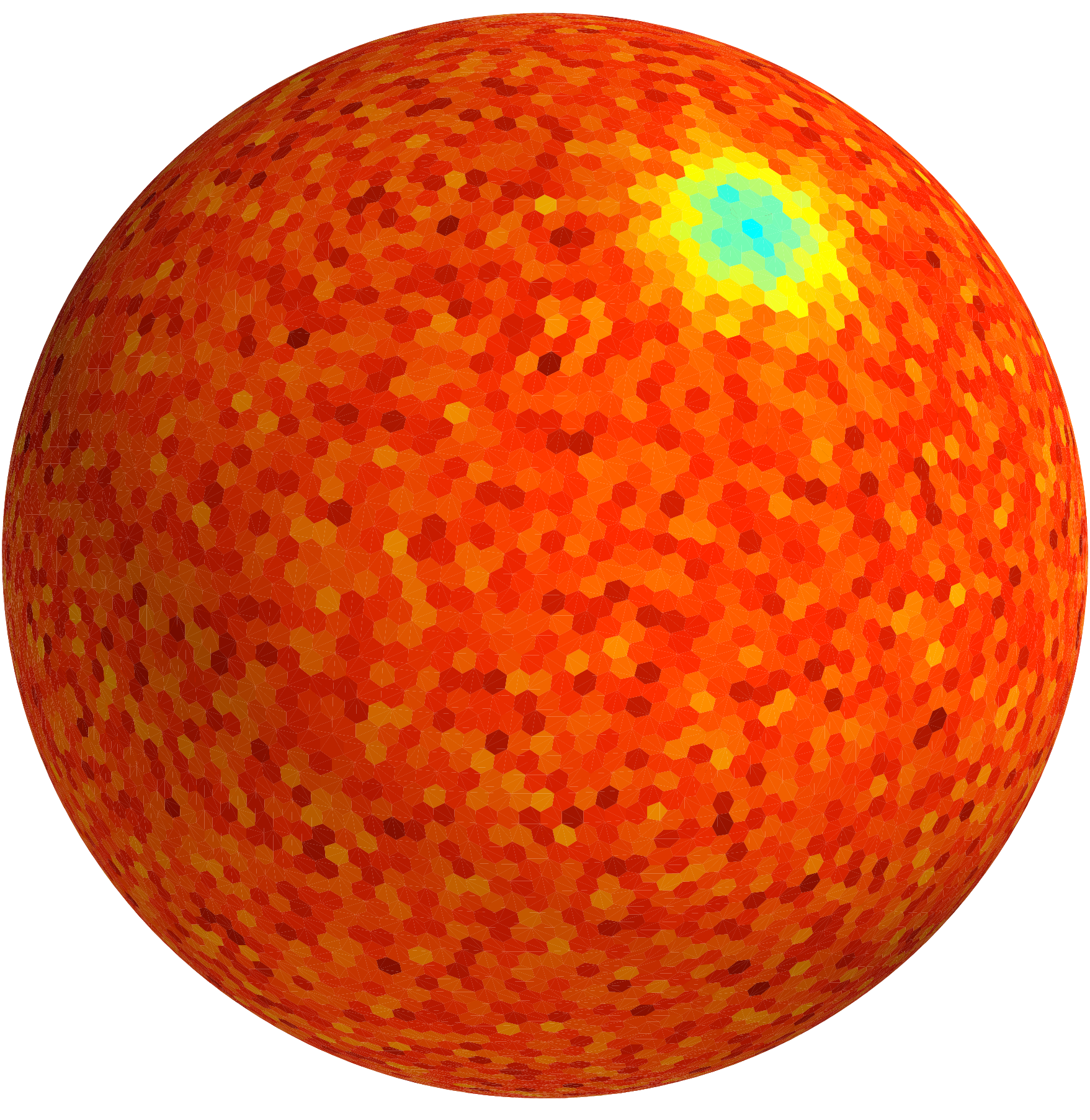}} &
  \resizebox{2.2cm}{!}{\includegraphics{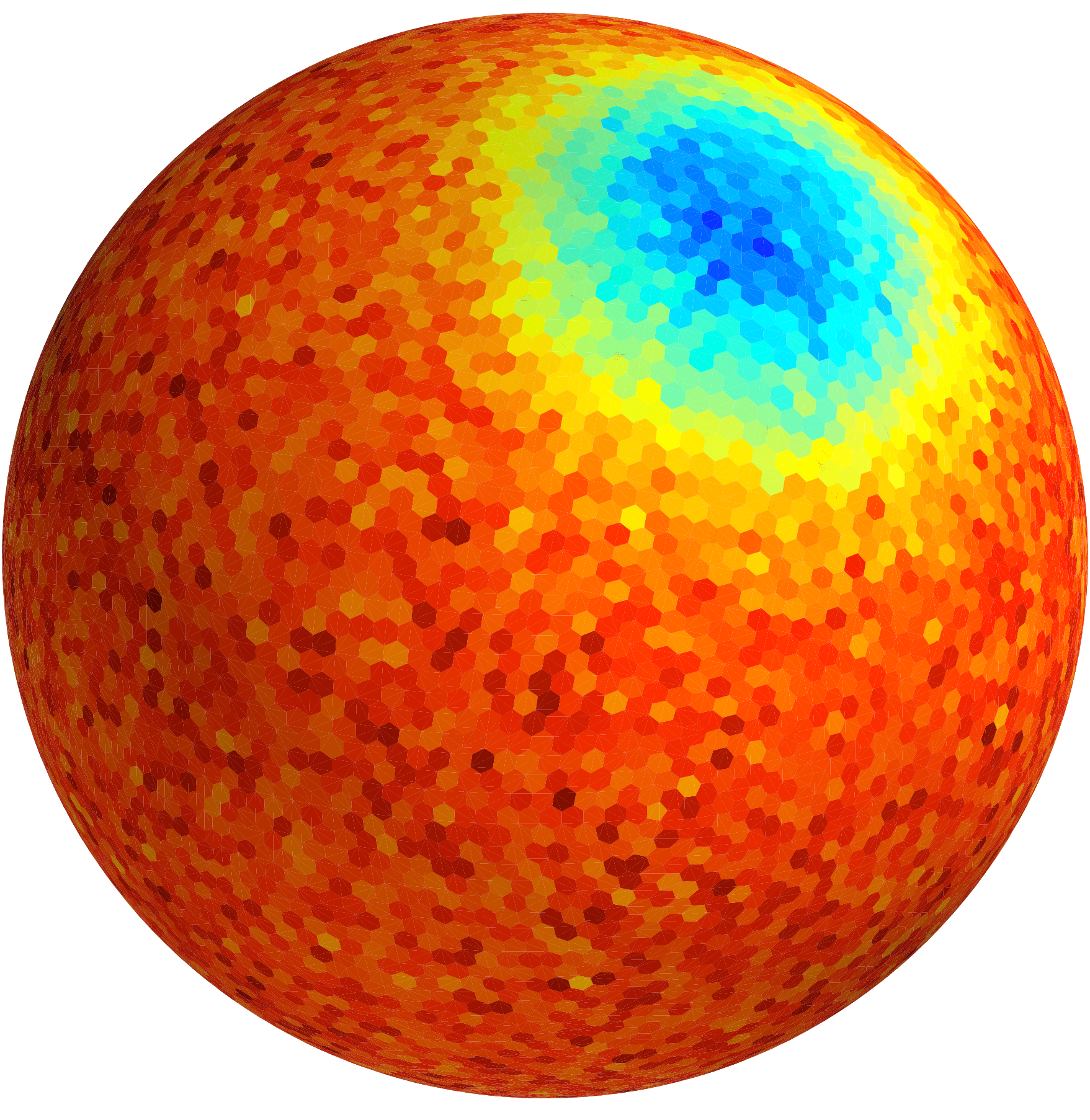}} &
  \resizebox{2.2cm}{!}{\includegraphics{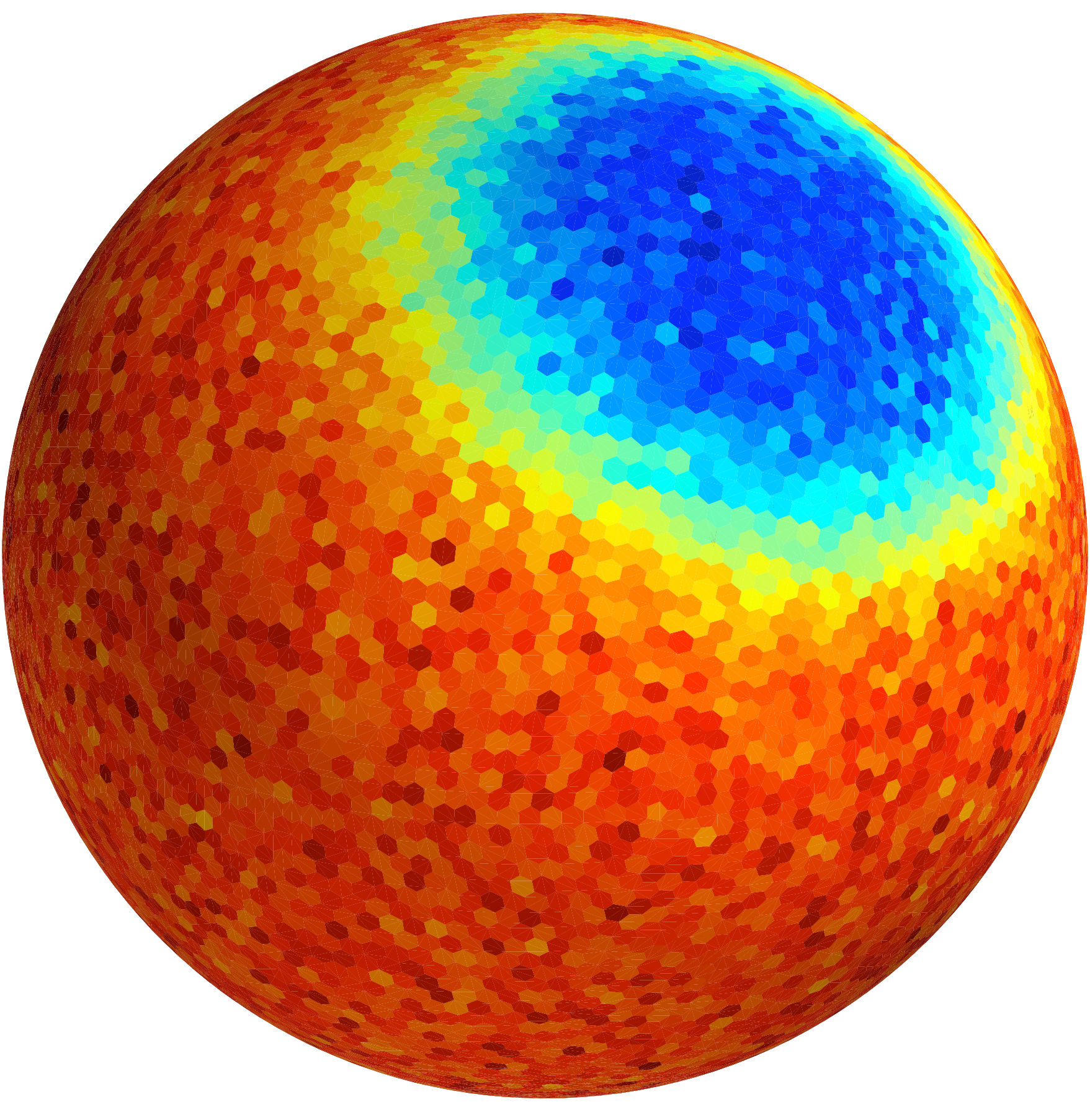}} \\
  $ $ & \textsf{0 min} & \textsf{4 min} & \textsf{10 min}
  \end{tabular}
\caption{(Color online) Growth of the PIP2-rich phase (blue {upper patch}).
The color scale shows the gradation of PIP2 content: {$\varphi$} is the relative concentration
difference between PIP3 and PIP2 at a given site. The system at initial time is in a uniform PIP3-rich phase, apart from an initial {PIP2-rich} seed germ of size $r_0$ larger than the threshold
radius $r_\mathrm{thr}$ (small circle.) After 4 minutes a PIP2{-rich} patch becomes
apparent and its radius saturates after approximately 10 minutes to the
equilibrium value $r_\mathrm{eq}$.} 
\label{snap}
\end{figure}

We represent the plasmamembrane by a lattice of $\mathcal{N}=10242$
({mostly }hexagonal) sites of area 
$\sigma \sim S / \mathcal{N}\sim(0.1\,\mu\mathrm{m})^2$ on a sphere surface $S=4 \pi R^2$ 
{with }radius $R=5\,\mu\mathrm{m}$ (Fig.~\ref{snap}). Each site is populated by a number of 
molecules of the chemical factors and their dynamics is described by standard master equations. 
For instance, the  PIP2 $\rightarrow$ PIP3 process is described by
\begin{eqnarray}
&&\!  \partial_t P (N_{\mathrm{PIP}_2}, N_{\mathrm{PIP}_3}, \ldots) = \nonumber\\
&+& 
  W ( \mathrm{PIP}_3 \rightarrow  \mathrm{PIP}_2) P (N_{ \mathrm{PIP}_2} - 1,
  N_{ \mathrm{PIP}_3} + 1, \ldots)  \nonumber\\
 &-& W ( \mathrm{PIP}_2 \rightarrow
   \mathrm{PIP}_3) P (N_{ \mathrm{PIP}_2}, N_{ \mathrm{PIP}_3}, \ldots),
\end{eqnarray}
where $P(N_\mathrm{X},...)$ is the probability to have at time $t$ a number
$N_\mathrm{X}$ of type $\mathrm{X}$ molecules {at }{a given} site (say, $i$.) 
The list of relevant reactions with their corresponding rates $W$ is given 
in Table~\ref{reac}~\footnote{
The values for processes involving cadherins are educated guesses since no
precise data are available. 
One order of magnitude changes in these values
do not result
however 
in appreciable modifications of the system dynamics.
}.
PIP2/3 diffusion is described by random jumps of a molecule from site $i$ to its neighboring site $j$ with rate $W(i\rightarrow j)=N_\mathrm{X} D /\sigma$, where 
$D=0.5 \, \mu\mathrm{m}^2/\mathrm{s}$ {is }{phospholipid diffusivity}{~\cite{GCT+05}}. 
Since {the diffusivity} of cytosolic enzymes (PI3K/PTEN)
is much larger than {that of} membrane pospholipids{~\cite{GCT+05}}, their 
distribution in the cytosol is treated as uniform.

For the simulations we use a variation of Gillespie algorithm~\cite{Gil77} taking into account the spatial non-uniformity of the system.
At time zero, a random number is generated to determine the next reaction or elementary diffusion process to occur, with a probability proportional to the corresponding $W$ factor from Table~\ref{reac}. Then, time is advanced as a Poisson process with rate again determined by the $W$ factors. These steps are repeated iteratively until the desired simulation time is reached. 

\begin{figure}
\centering
\includegraphics[width=8.cm]{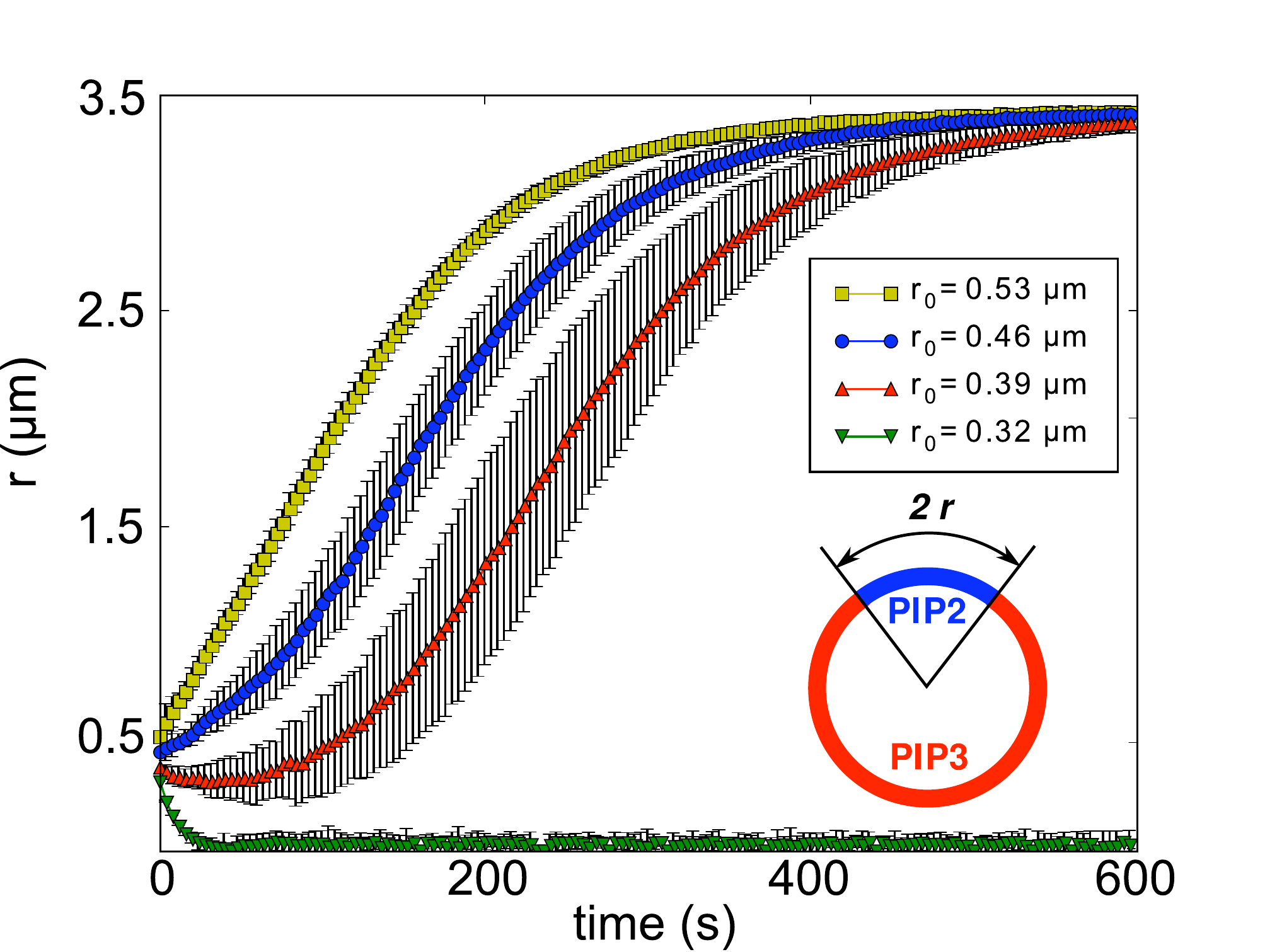}
\caption{(Color online) Growth of the PIP2-rich phase induced by a germ of initial radius $r_0$.
Germs with $r_0$ smaller than a threshold radius $r_\mathrm{thr}\sim 0.3\,\mu\mathrm{m}$ are
melted by diffusion, while larger germs grow to the equilibrium
value $r_\mathrm{eq}$. 
Error bars show standard deviations computed over $n=50$ different random
realizations of the process.
}
\label{radii}
\end{figure}

\begin{figure}
\centering
\includegraphics[width=8.cm]{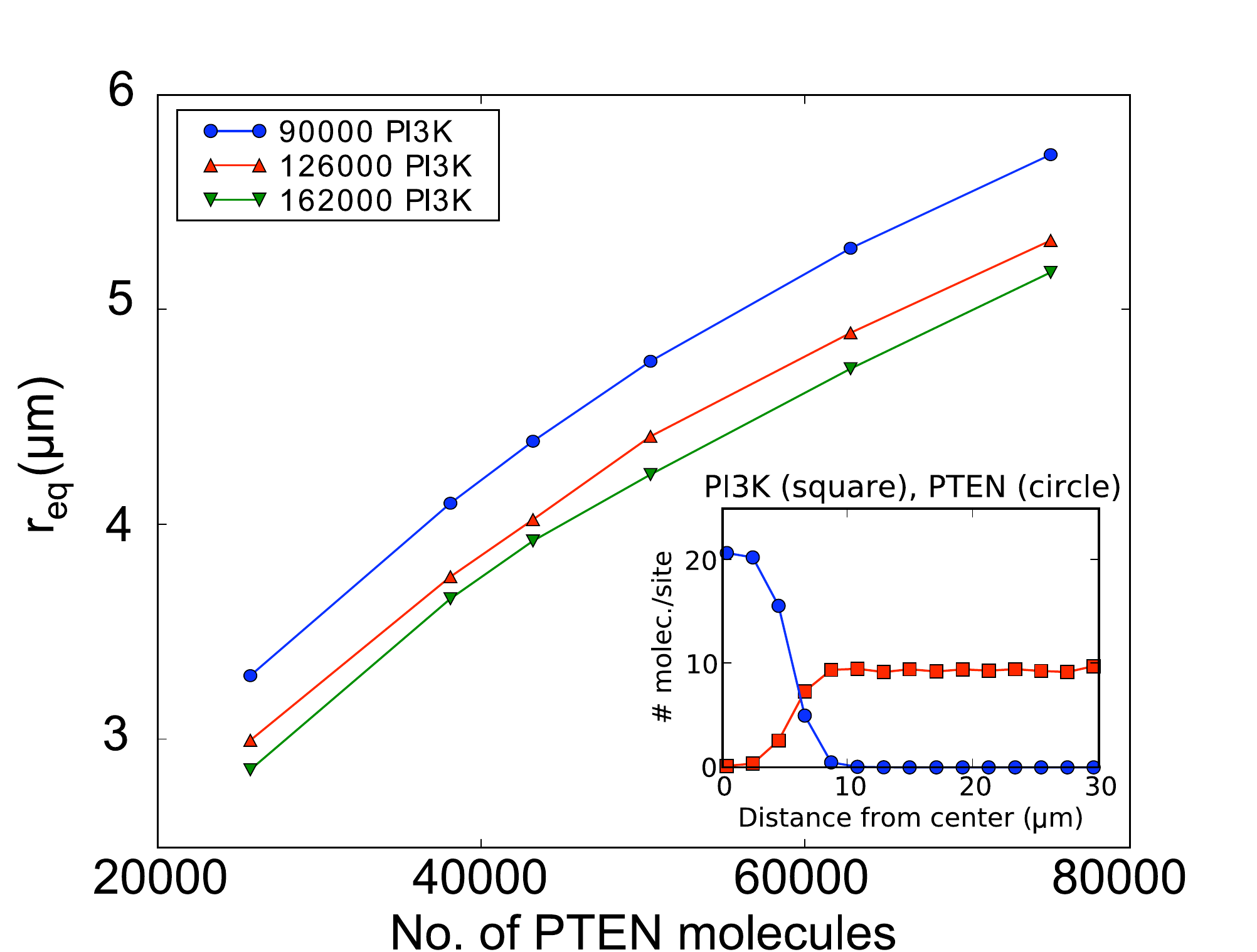}
\caption{(Color online) The equilibrium lumen size $r_\mathrm{eq}$ is an increasing function of
the relative number of PI3K and PTEN molecules. Inset: radial distribution
of PI3K (red {squares}, $N_\mathrm{PI3K}=1.26\cdot10^5$ molecules) and PTEN (blue {circles},
$N_\mathrm{PTEN}=0.43\cdot10^5$ molecules), from the patch center, at equilibrium.}
\label{pten}
\end{figure}

\begin{table}
\begin{tabular}{lll}
\hline
\hline
Reaction&\emph{W}&Rate{\ constants}\\
\hline
\footnotesize{$\mathrm{PIP2}\rightarrow\mathrm{PIP3}$} & \footnotesize{$
k_1 N_{\mathrm{PIP2}}N_{\mathrm{PTEN}}/(K+N_{\mathrm{PIP2}})$} & \footnotesize{$k_1=1,K=50$} \\
\footnotesize{$\mathrm{PIP3}\rightarrow\mathrm{PIP2}$} & \footnotesize{$
k_2 N_{\mathrm{PIP3}}N_{\mathrm{PI3K}}/(K+N_{\mathrm{PIP3}})$} & \footnotesize{$k_2=0.5,K=50$} \\
\footnotesize{$\mathrm{PTEN}\rightarrow\mathrm{PTEN}^{\ast}$} &
\footnotesize{$  k_3 N_{\mathrm{PTEN}}N_{\mathrm{PIP2}}$} & \footnotesize{$k_3=2\cdot 10^{-5}$} \\
\footnotesize{$\mathrm{PTEN}^{\ast}\rightarrow\mathrm{PTEN}$} &
\footnotesize{$  k_4 N_{\mathrm{PTEN}^{\ast}}$} & \footnotesize{$k_4=0.5$} \\
\footnotesize{$\mathrm{PI3K}\rightarrow\mathrm{PI3K}^{\ast}$} &
\footnotesize{$  k_5 N_{\mathrm{PI3K}}N_{\mathrm{Cad}^{\ast}}$} & \footnotesize{$k_5=2\cdot 10^{-5}$} \\
\footnotesize{$\mathrm{PI3K}^{\ast}\rightarrow\mathrm{PI3K}$} &
\footnotesize{$  k_6 N_{\mathrm{PI3K}^{\ast}}$} & \footnotesize{$k_6=0.1$} \\
\footnotesize{$\mathrm{Cad}\rightarrow\mathrm{Cad}^{\ast}$} &
\footnotesize{$  k_7 N_{\mathrm{Cad}}N_{\mathrm{PIP3}}$} & \footnotesize{$k_7=2\cdot 10^{-5}$} \\
\footnotesize{$\mathrm{Cad}^{\ast}\rightarrow\mathrm{Cad}$} &
\footnotesize{$  k_8 N_{\mathrm{Cad}^{\ast}}$} & \footnotesize{$k_8=0.5$} \\
\hline 
\hline
\end{tabular}
\caption{ List of chemical reactions involved in epithelial polarization and their corresponding rates. 
$\mathrm{X}^{\ast}$ denotes the membrane-bound, activated form of molecule $\mathrm{X}$. 
Rate constants $k$ are given in $\mathrm{s}^{-1}$. 
Activation rates can be transformed to $(\mathrm{s\,M})^{-1}$ units multiplying by 
Avogadro's number {$N_\mathrm{A}$.}
Michaelis-Menten constants $K$ are pure numbers and can be transformed to 
M units dividing by $N_\mathrm{A}/\cal{N}$. 
For the rate constants values see references in~\cite{GCT+05}.
}
\label{reac}
\end{table}
We suppose that {a circular PIP2-rich patch of radius $r_0$ is initially formed in the sea of
the PIP3-rich phase~\footnote{Where not otherwise stated, 
the following 
{experimentally}
realistic 
values
for the initial concentration are used:
$[\mathrm{PIP2}]+[\mathrm{PIP3}]=10^6$ {(according to~\cite{Van2001})},
$[\mathrm{PI3K}]=[\mathrm{Cad}]=10^5$ {(according to~\cite{Carpenter1990} and~\cite{Duguay2003}, respectively)}, $[\mathrm{PTEN}]=0.2\times 10^5$
(molecules/cell). {For the PTEN concentration, we assumed the same order of magnitude as in~\cite{GCT+05} and studied in Fig.~\ref{pten} the system behaviour on varying the PTEN molecules number}. The initial PIP3-rich phase is 98\% PIP3, 2\% PIP2.}}
and investigate its dynamics to check whether a stable polarization state is attained (Fig.~\ref{snap}). 
Fig.~\ref{radii} shows the time evolution of 
circular patches of different initial radii $r_0$. 
Patches smaller than a threshold radius $r_\mathrm{thr}\sim\,0.3\,\mu
\mathrm{m}$ are dissolved by diffusion and thermal processes, and do not
impair the stability of the uniform {PIP3-rich} phase. Conversely, patches
larger than $r_\mathrm{thr}$ grow in time triggering the separation of the
cell surface in a PIP2-rich and a PIP3-rich region 
and eventually reach an equilibrium{\ (Fig.~\ref{snap})}. Notably, the threshold radius 
{$r_\mathrm{thr}\sim\,0.3\,\mu \mathrm{m}$}
derived from the above calculation is consistent with the previously independently derived 
{value for the} adhesion radius, $r_\mathrm{a}$. 
The two phases are divided by an interface of characteristic width
$\delta\sim\sqrt{D/k_\mathrm{c}}\sim 1\,\mu\mathrm{m}$~\cite{GKL+07}, where
$k_\mathrm{c}$ is of the order of the catalytic constants of the 
two catalytic reactions of Table~\ref{reac}{\ (first two rows)}. 

The kinetic of this heterogeneous nucleation process can be understood 
in terms of non-equilibrium, reaction-diffusion stochastic dynamics. 
In reaction-diffusion systems instabilities are often produced by Turing's mechanism~\cite{Tur52}. 
Here we find that pattern formation starting from a locally stable homogeneous state is triggered by a local perturbation by a 
nucleation center of size $r_0$ larger than a critical size $r_c$~\cite{Bra94,Sch00,GKL+07,GKL+08}.

Fig.~\ref{pten} shows that the equilibrium size {$r_{eq}$} of the PIP2{-rich} patch, and therefore of the lumen, is controlled by the number of PI3K and PTEN molecules.
In the absence of any limiting mechanism, the growth of the
PIP2{-rich} patch would {in fact} lead to {a PIP2-rich} phase
completely invading the cell surface. However, due to the coupling to a
finite PTEN {and PI3K} reservoir, the system self-tunes to a
phase-coexistence state and the process stops when the PIP2{-rich} patch reaches 
{the equilibrium size $r_{eq}$}~\cite{GCT+05,GKL+07,FCG+08}. 
{Interestingly, the fact that the size of the PIP2-rich patch, and consequently of the
lumen, is controlled by the precise number of PTEN molecules, is in
qualitative agreement with the observation that deletion of a single PTEN allele can
interfere with the polarization process~ \cite{GYB+06}}.

The observation in the present scenario of a comparatively large threshold radius, of the order of one tenth of the cell size, suggests the existence of a correspondingly large 
barrier {of effective energy} dividing the uniform state from the phase-separated one.
This prevents thermal 
{and chemical noise} from triggering spontaneous symmetry breaking and lumen formation.
However, an external 
mechanical
action creating a sizeable PIP2{-rich} patch, 
{due to
the presence of
localized regions of high membrane-curvature}, can overcome the barrier
and start polarization~\footnote{{Interestingly, the role of
mechanical forces has been suggested also in 
other settings of tissue
morphogenesis~\cite{Shr05}.}}. 
Our picture also explains tight junction localization. 
Experimental data show that the stable binding of tight junction proteins 
to the membrane requires both a 
protein complex named PAR3-PAR6, 
which is localized in the PIP2-rich phase by a
chain of reactions, \emph{and} cell-cell contacts, which are maintained
only in the PIP3-rich phase~\cite{MB05}. 
The spontaneous aggregation of tight junctions is thus constrained 
by a biochemical logical AND
to take place only on the 
ring-shaped boundary separating the PIP2-rich from the PIP3-rich phase.

\section{Mean-field}\label{sec:mf}
In this Section we show that the results of the simulations can be
conveniently interpreted in terms of an effective mean-field model,
following the approach detailed in Refs.~\cite{GKL+07,GKL+08}.

Fast diffusion of PI3K and PTEN enzymes in the cytosol, and
the conservation law 
$\mbox{[PIP2]}+\mbox{[PIP3]}=c$      
allow to
effectively describe the state of the cell membrane in terms of the
configuration of the single-component concentration
field~\cite{GKL+07,GKL+08}:
\begin{equation}
\varphi=\mbox{[PIP3]}-\mbox{[PIP2]}.
\end{equation}
The resulting effective equation for $\varphi$ can be set in the simple
Landau-Ginzburg form:
\begin{equation}
\partial_t \varphi = D \nabla^2 \varphi + V' (\varphi) + \xi,
\label{lg}
\end{equation}
complemented by an integral costraint expressing the coupling of the
concentration field $\varphi$ to the reservoir of free cytsolic enzymes (see Ref.s~\cite{GKL+07,GKL+08} and Supplementary Information in~\cite{GCT+05}.)
In Eq.~(\ref{lg}), 
$D$ is the diffusivity of lipids on the cell membrane, $V(\varphi)$ is an effective potential, and $\xi$ is a stochastic term taking into account the effect of thermal and chemical noise.

The mean-field effective potential $V(\varphi)$ can be easily derived,
via a quadratic approximation,
from the stochastic model described in Sect.~\ref{sec:simul} under the
assumption that the cytosolic PI3K, PTEN and Cad
fields are in approximate equilibrium with the membrane PIP2 and PIP3
fields, and therefore ``slaved" to the $\varphi$ field~\cite{GKL+07,GKL+08}:
\begin{eqnarray}\label{f}
V'(\varphi)&=& - \alpha\ \frac{c^2 - \varphi^2}{2 K + c + \varphi}  + \alpha'\ \frac{ c^2 - \varphi^2}{2 K + c - \varphi},
\end{eqnarray}
where
\begin{eqnarray}
\alpha&=&\frac{k_2\,k_3}{k_4}\,[\mathrm{PTEN}]_\mathrm{free},\label{alpha}\\
\alpha'&=&\frac{k_1\,k_5}{k_6}\,[\mathrm{PI3K}]_\mathrm{free}\cdot\frac{k_7}{k_8}\,[\mathrm{Cad}]_\mathrm{free}.\label{alpha1}
\end{eqnarray}
The terms in the r.h.s. of Eq.~(\ref{f}) describe respectively conversion of PIP3 into PIP2 due to the action of PTEN, 
and conversion of PIP2 into PIP3 due to the action of PI3K activated by
cadherins (Fig.~\ref{net}). 
The quadratic terms $\propto c^2 -
  \varphi^2$ encode respectively the PIP2 $\rightarrow$PTEN and the PIP3$\rightarrow$Cad$\rightarrow$PI3K feedback loops~(Fig.~\ref{net}). 
In particular, $\alpha'=0$ when cadherin links are broken.
 
\begin{figure}
\centering
\includegraphics[width=9.cm]{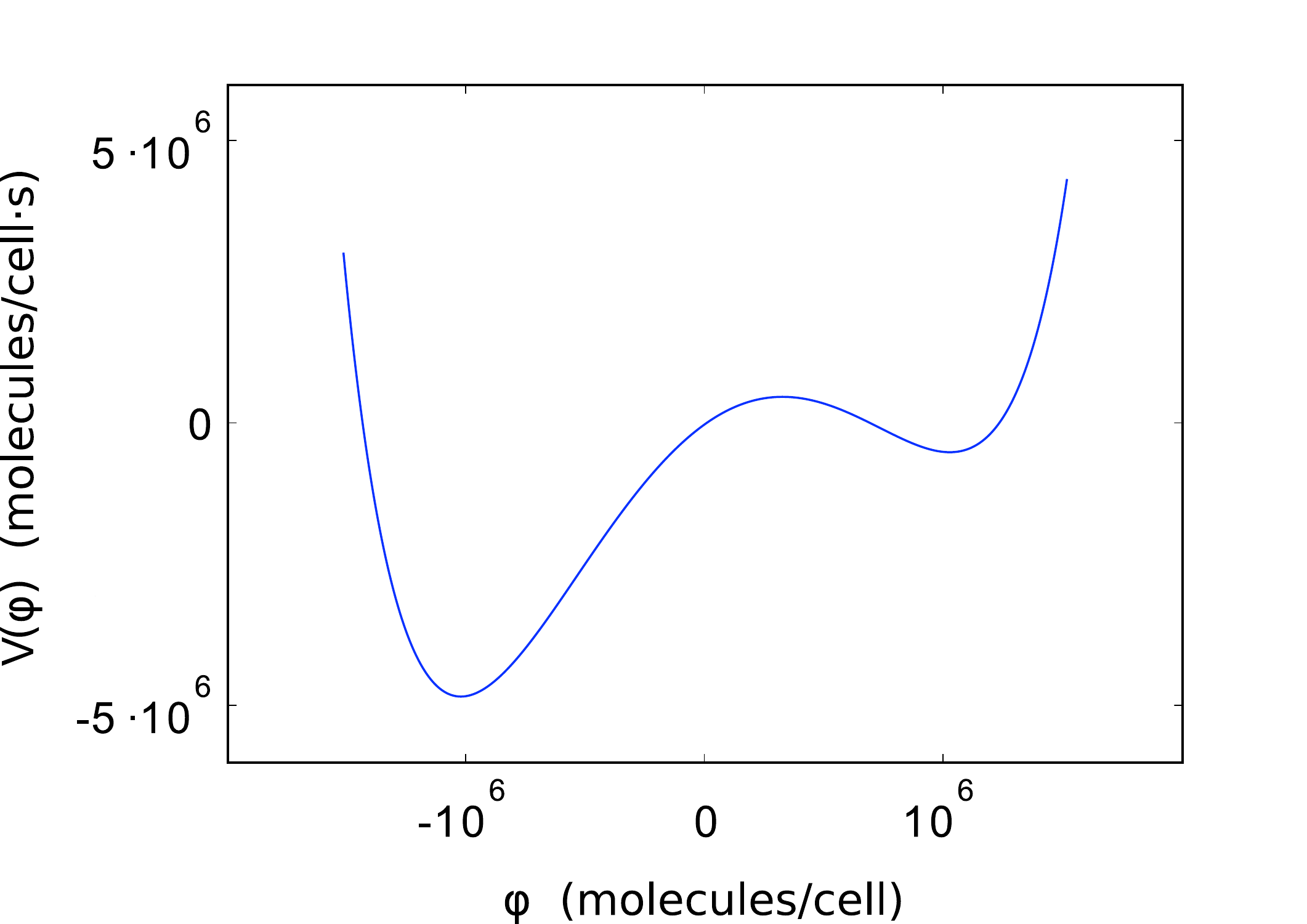}
\caption{(Color online) Graph of the effective potential $V(\varphi)$ defined by
(\ref{f},\ref{alpha},\ref{alpha1}), where the cytosolic values
$[\mathrm{PTEN}]_\mathrm{free}\simeq 1.1\cdot 10^4$,
$[\mathrm{PI3K}]_\mathrm{free}\simeq 3.7\cdot 10^4$,
$[\mathrm{Cad}]_\mathrm{free}\simeq 5.4\cdot 10^3$ (molecules/cell)
were computed from simulations of the stochastic model described in 
Sect.~\ref{sec:simul} at time $t=30~\mathrm{s}$, after the fast equilibration 
of the corresponding cytosolic pools of enzymatic factors
with the initial configuration of the $\varphi$ field. 
The effective potential $V(\varphi)$ 
has two minima, the left hand one corresponding to a stable PIP2-rich
and the right hand one corresponding to
a metastable PIP3-rich phase.
The two phases are separated by an effective energy barrier. 
}
\label{x}
\end{figure}

In a wide region of parameter space around the realistic parameter values
from Table~\ref{reac}, the effective potential $V(\varphi)$ is bistable (Fig.~\ref{x}; 
for a detailed description of the bistability region, see Ref.~\cite{GNS+09}).
The two potential wells in Fig.~\ref{x} correspond to a stable PIP2-rich
and a metastable PIP3-rich phase, separated by an energy barrier $\Delta V$. 

The mean-field model~(\ref{lg},\ref{f}) 
and the bistability of the effective potential $V(\varphi)$
provide an interpretation to the simulation results, 
showing that the stable polarized state characterized by the coexistence
of PIP2 and PIP3 in complementary regions is separated 
from the metastable PIP3-rich phase by an effective energy barrier $\Delta V$. 
According to the theory of Landau-Ginzburg equation, PIP2-rich seeds
larger than a critical value are bound to 
expand in the PIP3-rich sea with a velocity proportional to $1/\Delta V$~\cite{Bra94}. 

The cytosolic concentrations $\mathrm{[PI3K]}_{\mathrm{free}}$,
$\mathrm{[PTEN]}_{\mathrm{free}}$, and
$\mathrm{[Cad]}_{\mathrm{free}}$ 
appearing in~(\ref{f}) may be expressed as integrals of the concentration
field $\varphi$~\cite{GKL+08,GNS+09}.
The resulting global coupling has the effect of driving dynamically the
cell membrane towards an equilibrium polarized state where the PIP2-rich and PIP3-rich phases coexist: the growth of the PIP2-rich phase ``eats up" free PTEN molecules from the cytosol, decreasing $\Delta V$ until phase coexistence is reached~\cite{GKL+08, GNS+09}. 
This process may be understood via a simple physical analogy
with the non-equilibrium process taking place during the liquid-vapor transition
in a sealed vessel: there,
the rise of the vapor pressure (which in our analogy corresponds to the number
of cytosolic enzymes) provides 
a negative feedback, slowing down the growth 
of the vapor phase and eventually leading the system to a state of phase
coexistence.
The main difference between the two systems is that in the liquid-vapor
transition a local conservation law holds for the particle field, while in
the growth of signaling domains on the cell membrane the $\varphi$ field
satisfies only an approximate \emph{global} constraint~\cite{GKL+08}
encoded in the integral expressions for the coefficients $\alpha,\,\alpha'$
appearing in
(\ref{f},\ref{alpha},\ref{alpha1}).

It is worth observing here that while spontaneous polarization in eukaryotic
directional sensing~\cite{GCT+05} can be described in terms of an homogeneous nucleation
process, whereby seeds of a PIP3-rich phase are created by thermal and
chemical noise in the sea of the PIP2-rich phase and grow by a coarsening
process~\cite{GKL+07,GKL+08}, the present picture of
epithelial polarization reminds instead \emph{heterogeneous} nucleation,
\emph{i.e.} a situation where the effective potential barrier $\Delta V$
is so high that spontaneous nucleation does not occur in typical
observational times, and needs to be triggered by the introduction of a large
enough nucleation germ.

\section{Discussion}\label{sec:disc}
We have shown that a simple symmetry breaking mechanism, informed with the recently discovered  
biochemical and biophysical details of the system, accounts for a wealth of 
morphogenetic processes in  epithelial polarization. The model makes specific predictions on the dependence 
of the threshold radius, lumen size, tight junction positioning and width, on the biochemical system parameters. 
{Our results shed light on the role of PTEN as a tumor suppressor
protein, whose expression levels are known to be critical to prevent the onset of cancer.}
In particular, our models predicts that by decreasing the number of PTEN
molecules, the lumen size should decrease, and for very low PTEN levels no lumen at all
should form.
The experimental validation of the model could be performed by genetic manipulation of the amounts or activity of cadherins, PI3K and PTEN. 

{We have also shown that curvature induced forces are a very plausible candidate for triggering the symmetry breaking process at the right time. 
This could be verified experimentally by trying to induce localized formation of a growing PIP2 patch and lumen by mechanically 
breaking adhesion bonds in localized regions of the membrane of epithelial cells surrounded by extracellular matrix.}
Under these conditions, our model predicts that only breaking adhesion
bonds in regions larger than $r_\mathrm{thr}$ should induce the formation
of a growing patch of the PIP2-rich phase, while smaller PIP2-rich
patches, induced by breaking adhesive bonds on smaller regions, should
shrink spontaneously.

Interestingly, 
the bistable {PI3K-PTEN} 
module {here described plays} also a key role in chemotaxis, where PI3K is {initially} activated by chemotactic receptors {(see the review~\cite{Iglesias2008} and ref.s therein)}  rather than by {adhesive} receptors. 
While {experimental} evidences and our results suggest that epithelial
polarization is induced by a nucleation center of the PIP2-rich phase
generated by mechanical forces, the polarization of migrating cells 
{is likely to be} triggered by spontaneous fluctuations in PIP3{/PI3K} 
levels~\cite{GKL+07}. 

{PIP3 localization 
also
regulates chemotactic polarization~\cite{GCT+05, GKL+07} and
cell spindle 
orientation~\cite{TMM+07}}. 
The similarity underlying the mechanisms in these very different
aspects of cell life hints to the possibility that phase separation 
phenomena might have a general role in the
cell~\cite{GCT+05, GKL+07} and in its nucleus~\cite{xci_nicodemi,nic_genetics}. 
The principles emerging here could explain {in a universal way} the deep
analogies observed in a variety of cellular processes involving spatial
polarity formation~\cite{Comer07}.\\

\section*{ACKNOWLEDGMENTS}
We thank G. Boffetta for discussions, hospitality at ISAC-CNR and access to
computational facilities, and S. Vegetti for suggestions.
This work was partially supported by Telethon-Italy GGP04127,
AIRC, MIUR (PRIN 2007BMZ8WA), Regione Piemonte, PRESTO, Fondazione CRT,
Ministero della Salute.

\end{document}